\documentclass[pra,reprint, nofootinbib,superscriptaddress]{revtex4-1}
\usepackage{graphicx}
\usepackage[colorlinks=true,linkcolor=blue,citecolor=red,urlcolor=blue]{hyperref}
\usepackage{amsmath}
\usepackage{bm}
\usepackage{soul}
\usepackage{subfigure}
\usepackage{color}
\usepackage{cancel}
\usepackage{comment}
\usepackage{physics}
\usepackage{soul,xcolor}

\newcommand{\be}{\begin{equation}}
\newcommand{\ee}{\end{equation}}

\begin{document}
\title{Stationary States and Instabilities of a M\"{o}bius Fibre Resonator} 
\author{Calum Maitland}
\email{cm350@hw.ac.uk}
\affiliation{Institute of Photonics and Quantum Sciences, Heriot-Watt University, Edinburgh EH14 4AS, UK} 
\affiliation{School of Physics and Astronomy, University of Glasgow, Glasgow G12 8QQ, UK}
\author{Matteo Conforti}
\affiliation{Universit\'{e} Lille, CNRS, UMR 8523-PhLAM-Physique des Lasers Atomes et Mol\'{e}cules, F-59000 Lille, France} 
\author{Arnaud Mussot}
\affiliation{Universit\'{e} Lille, CNRS, UMR 8523-PhLAM-Physique des Lasers Atomes et Mol\'{e}cules, F-59000 Lille, France} 
\affiliation{Institut Universitaire de France (IUF)}
\author{Fabio Biancalana}
\affiliation{Institute of Photonics and Quantum Sciences, Heriot-Watt University, Edinburgh EH14 4AS, UK}

\begin{abstract}

We examine the steady state and dynamic behaviour of an optical resonator comprised of two interlinked fibre loops sharing a common pump. A coupled Ikeda map  models with great accuracy the field evolution within and exchange between both fibres over a single roundtrip. We find this supports a range of rich multi-dimensional bistability in the continuous wave regime, as well as previously unseen cavity soliton states. Floquet analysis reveals that modulation and parametric instabilities occur over wider domains than in single-fibre resonators, which can be tailored by controlling the relative dispersion and resonance frequencies of the two fibre loops. Parametric instability gives birth to train of pulses with a peculiar period-doubling behavior.

\end{abstract}

\date{\today}
\maketitle

\section{Introduction}

Optical resonators are complex physical platforms, exhibiting an even richer range of phenomena than single-pass nonlinear systems due to their driven-dissipative nature. They are of increasing importance in metrology as sources of highly tunable, broadband frequency combs \cite{DelHaye2007, Kippenberg2011, Schliesser2012}. The Kerr cavity soliton is one of the fundamental states responsible for generating these combs \cite{Leo2013, Coen2013a, Herr2014, Hansson2015, Bitha2019}. Further, they have been proposed as cryptographic tools due to their chaotic output  \cite{Ramos2000, Imai2009, Tunsiri2012}.
Nonlinear optical resonators can be precisely modelled by the so-called Ikeda map \cite{Ikeda1979, Steinmeyer1995}, which describes separately the evolution of the electric field as it propagates through the cavity, and the boundary conditions which account for the injection of pump light and transmission of the cavity field between each round-trip. 

\begin{figure}[h]  
\centering
\includegraphics[width=0.95\linewidth]{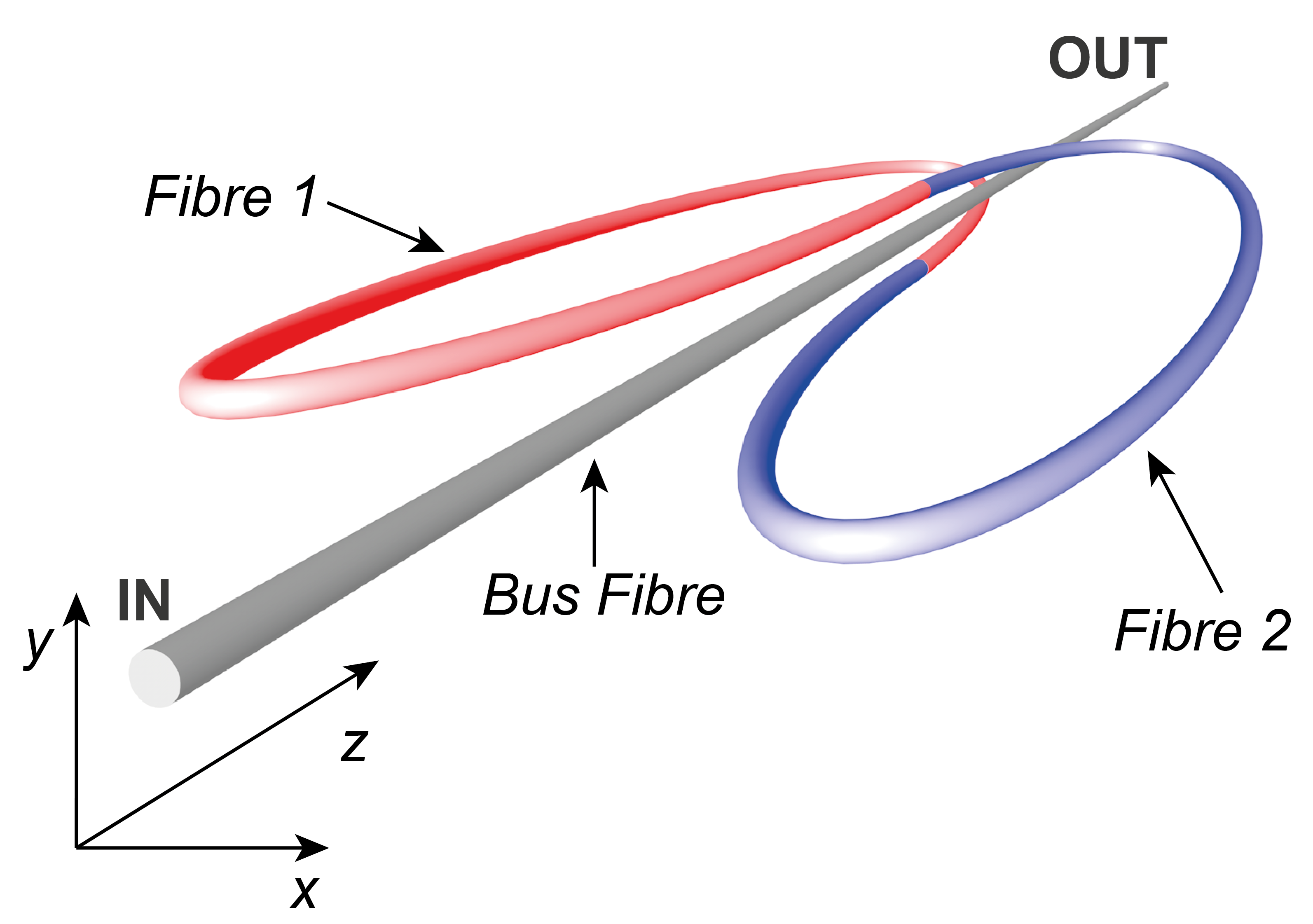}
\caption{\small{Schematic of the M\"{o}bius fibre loop resonator. A pair of optical fibres are linked as shown, such that the output of each is fed into the other. Both fibre inputs are coupled evanescently to a bus fibre, which carries pump light into each and allows for partial transmission of each fibre's output.}}  
\label{fig: ResonatorSketch}
\end{figure}

In this work, we propose a resonator composed of two fiber loops sharing a common pump. This geometry is reminiscent of the  M\"obius strip. Indeed, the two fiber coils are not closed, but they form an unique loop, in the same way as a M\"obius surface has an unique side. The structure of the M\"obius fibre resonator we consider is shown in figure \ref{fig: ResonatorSketch}. We model the light propagation inside this resonator by means of a coupled Ikeda map. Several works have considered extended/multi-dimensional Ikeda maps and mean-field approximations by Lugiato-Lefever equations with nonlinear coupling between the fields of a single resonator  \cite{Daguanno2016, DAguanno2017, Yi2017, Guo2017, Woodley2018,Haelterman1992,Haelterman1992a,Haelterman1993}, but to our knowledge no study has addressed the scenario of a pair of resonators in which the output of one is fed back into the other and vice versa.

The dimensionless Ikeda map connecting the fibre fields between round-trips (labelled with an integer $m$) is 
\begin{equation} \label{eq: normalisedIkeda}
\begin{split}
A_1^{m+1}(z=0, t) = \sqrt{\theta} A_{in} + \sqrt{1-\theta} e^{-i \delta_2} A_2^m(z=1, t)\\
A_2^{m+1}(z=0, t) = \sqrt{\theta} A_{in} + \sqrt{1-\theta} e^{-i \delta_1} A_1^m(z=1, t)\\
\end{split}
\end{equation}
while intra-fibre propagation is described by a lossy nonlinear Schr\"{o}dinger equation (NLSE)
\begin{equation} \label{eq: normalisedNLSE}
i \partial_z A_n^m = \eta_n {\partial_t}^2 A_n^m + i \beta_{3,n} {\partial_t}^3 A_n^m - {|A_n^m|}^2 A_n^m -i \frac{\alpha_{i}}{2} A_n^m.
\end{equation}
for fibres indexed as $n=1,2$. We work in dimensionless units in which the intra-fibre propagation coordinate $z = \tilde{z}/L$ is scaled to the fibre length $L$ (denoting here and subsequently all physical counterparts of dimensionless quantities with a tilde $\sim$). Time $t = \tilde{t} \sqrt{2/|\beta_{2,2}| L}$ is scaled by the second-order dispersion coefficient of the second fibre  $\beta_{2,2} \equiv [\partial_\omega^2 \beta_2]_{\omega_{0}}$, given propagation constants $\beta_2(\omega)$ in the second fibre and pump frequency $\omega_0$. Hence $\eta_2 \equiv \text{sgn}(\beta_{2, 2})$ and $\eta_1 = \text{sgn}(\beta_{2, 1})|\beta_{2, 1} / \beta_{2, 2}|$ and the dimensionless third order dispersion parameters are related to its physical counterpart $\beta_{3, n} = \sqrt{2/{|\beta_{2,2}|}^3 L}  [\partial_\omega^3 \beta_n]_{\omega_{0}}/3$. The fibre fields $A_n^m$ are related to physical electric fields as $A_n^m=\sqrt{\gamma L} \tilde{A}_n^m$ where $\gamma$ is the Kerr nonlinear coefficient, and similarly the pump field $A_{in} = \sqrt{\gamma L}  \tilde{A}_{in}$. Since we work with a pump which is implicitly a time-independent constant, our frequency variable $\Omega$ expresses a scaled detuning of the physical frequency $\omega$ from $\omega_0$, that is $\Omega = (\omega-\omega_0)\sqrt{|\beta_{2,2}| L/2}$. $\theta$ parametrises the transmission of pump light into the fibres and fibre output fields into the bus waveguide. The absorption coefficient is also scaled by the fibre length $\alpha_i = \tilde{\alpha}_i L$, while 
the detunings $\delta_n=2k\pi-\beta_n(\omega_0)L$ measure the phase difference accumulated per round-trip with respect to the nearest single-loop resonance indexed by the integer $k$.

In this article, we investigate the power output of the M\"{o}bius resonator in different dynamical situations. First we examine steady states in the continuous-wave (time independent) limit, where the output from either fibre is independent of both the round trip number $n$ and the intracavity time coordinate $t$. This reveals an extended set of solutions exhibiting bistability, whose symmetry depends on the relative detuning of the two fibre loops. In the following section we consider the dynamical case in which the output varies over time, with a period corresponding to an integer number of round-trips. Here, allowing the fibre loops to have different dispersive characteristics gives rise to a new class of Kerr cavity solitons. The structure of these solitons is sensitive to the relative second and third order dispersion coefficients in each fibre. Finally we derive the modulation instability (MI) spectrum by applying Floquet analysis, which demonstrates the existence of additional Arnold instability tongues beyond those found in single fibre resonators, arising from dispersion variations between the two loops. 

\section{Homogeneous Steady States}

\begin{figure}[h] 
\centering
\includegraphics[width=0.95\linewidth]{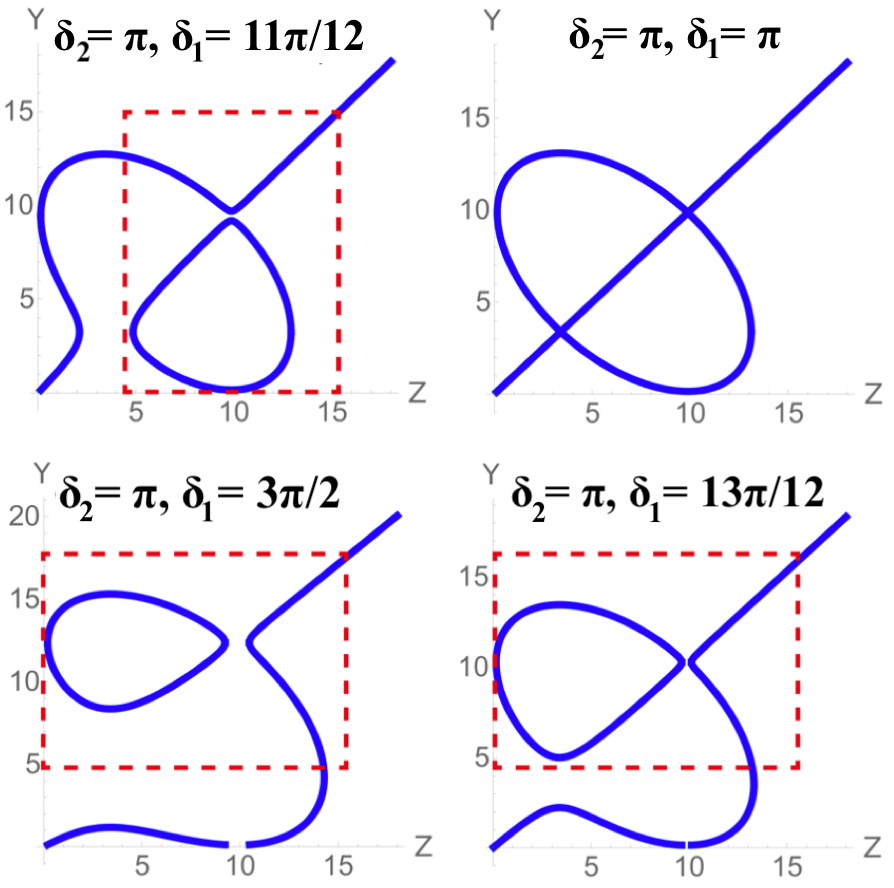}
\caption{\small{Solutions to Eq.~\eqref{eq: pumpindSS} showing the possible stationary state powers $Y, Z$ in the first and second fibre respectively, given fixed $\delta_2=\pi$ and different values of $\delta_1$ (increasing going clockwise from top-left). The solutions approximate sections of elliptic curves in the regions enclosed by the dashed red boxes. Here $\theta=2/15$ and $\alpha_i=1$.}}  
\label{fig: TempSS}
\end{figure}

\begin{figure}[h] 
\centering
\includegraphics[width=0.8\linewidth]{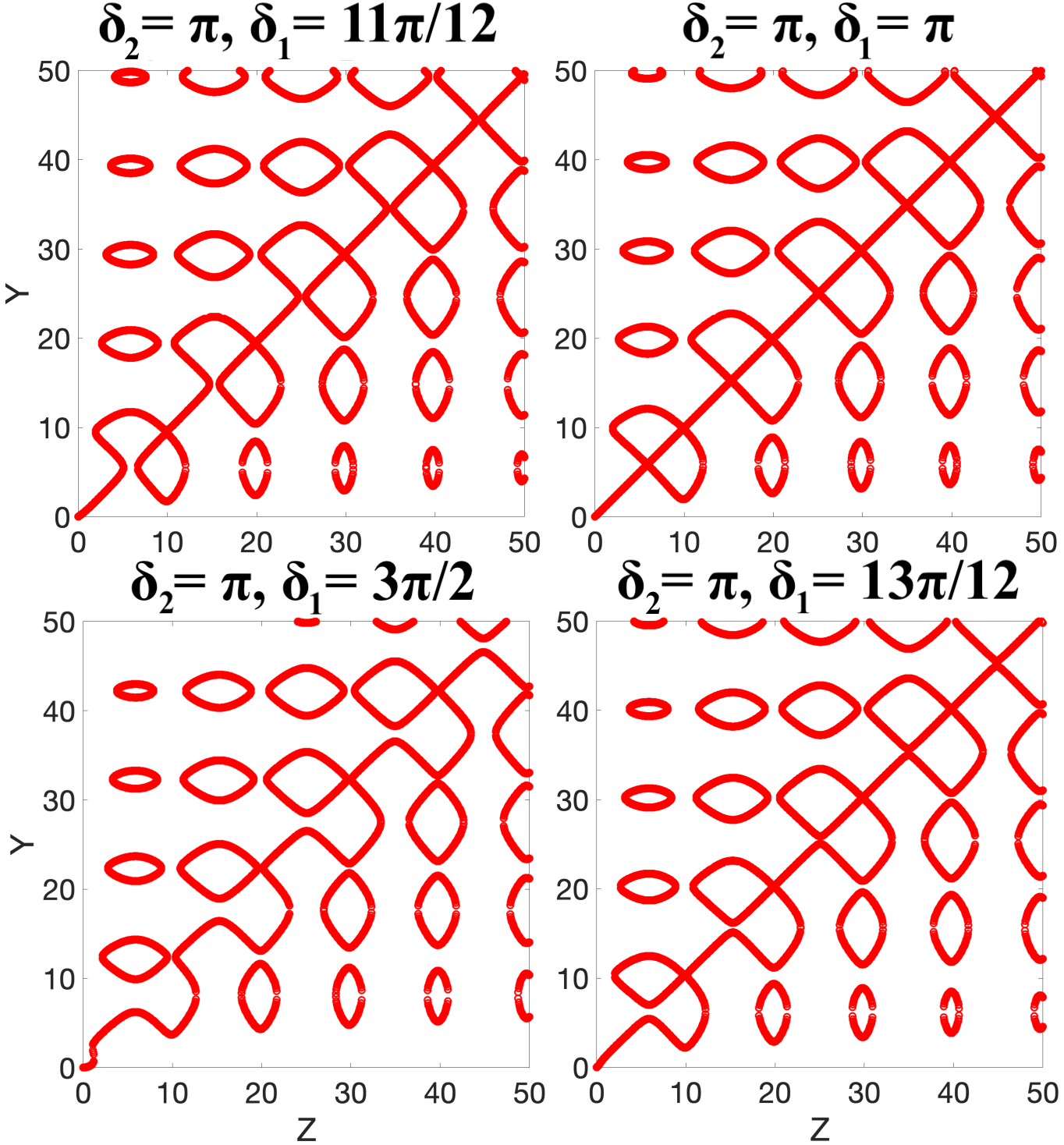}
\caption{\small{Solutions to Eq.~\eqref{eq: exactpumpindSS} showing the possible stationary state powers $Y, Z$ in the first and second fibre respectively, given fixed $\delta_2=\pi$ in the  and different values of $\delta_1$ (increasing going clockwise from top-left). Here $\theta=2/15$ and $\alpha_i=1$.}}  
\label{fig: exactSS}
\end{figure}

\begin{figure}[h] 
\centering
\includegraphics[width=0.8\linewidth]{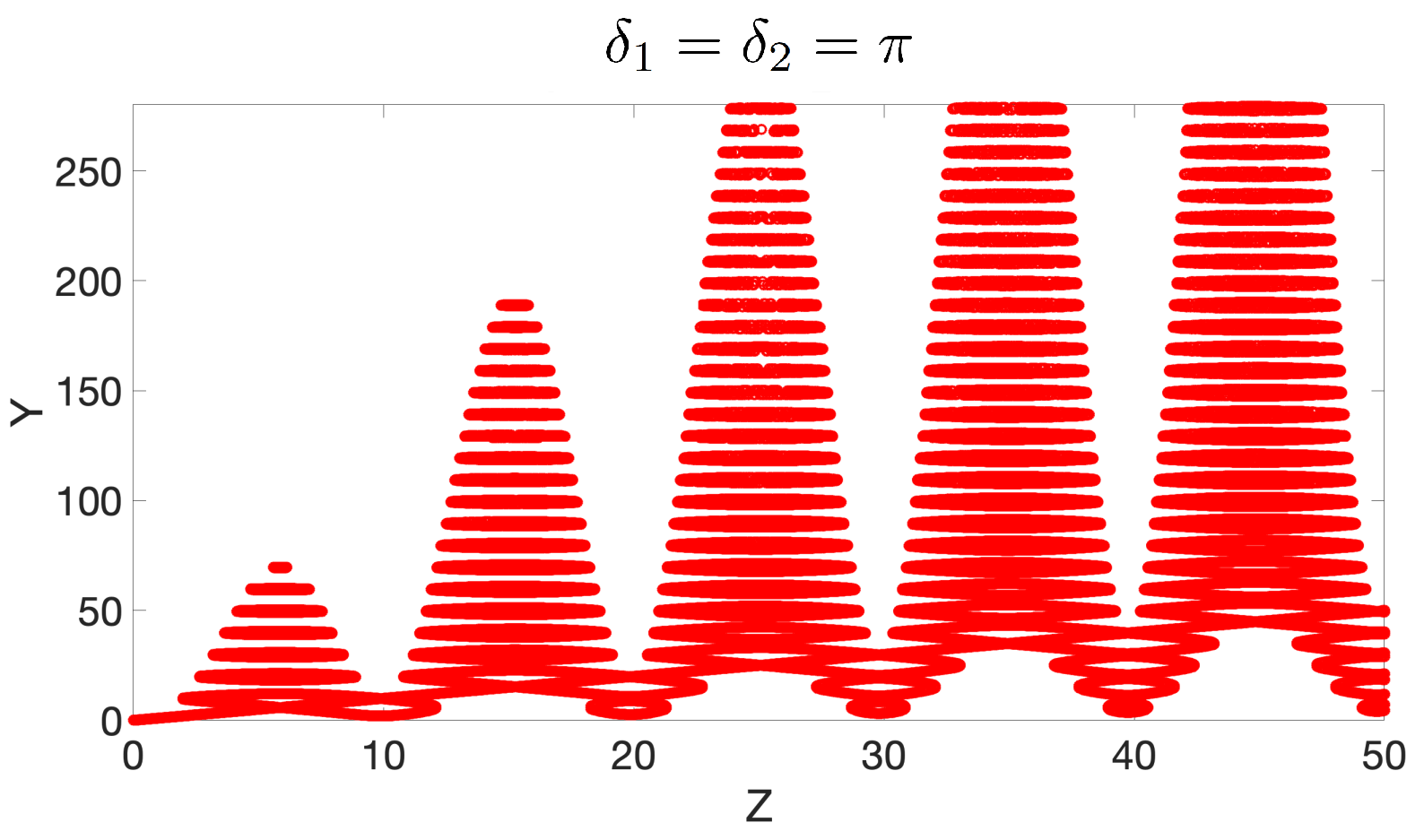}
\caption{\small{The same set of solutions to Eq.~\eqref{eq: exactpumpindSS} from figure \ref{fig: exactSS} with detuning $\delta_1=\delta_2=\pi$ plotted with an an extended $Y$ axis. A ladder-like structure of bands of states exists; typically bands except the lowest one closest to $(Y,Z)=(0,0)$ are unstable. }}  
\label{fig: exactSS2}
\end{figure}

\begin{figure}[h] 
\centering
\includegraphics[width=0.95\linewidth]{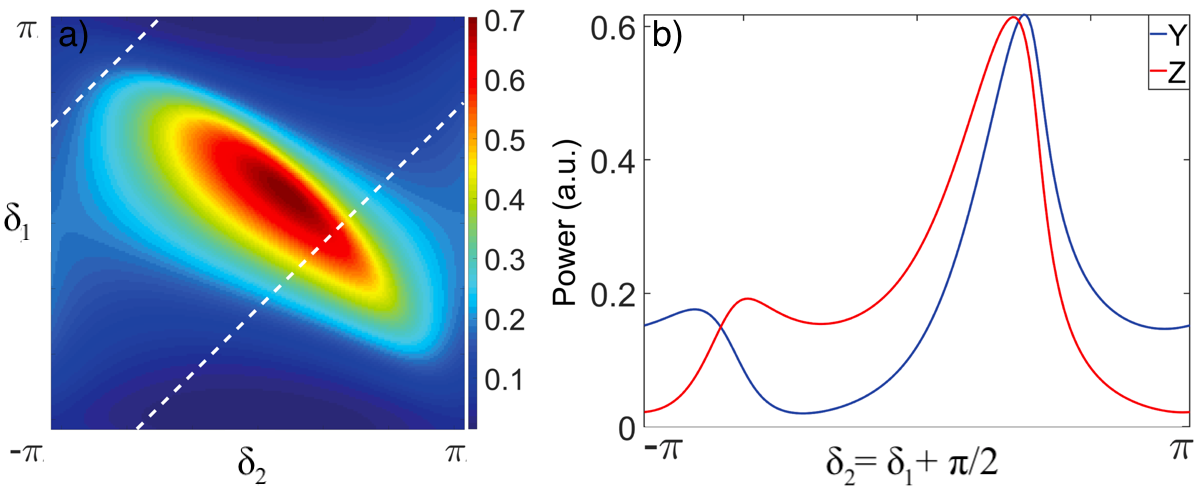}
\caption{\small{\textbf{(a)} Homogeneous steady-state power in the first fibre as a function of both fibre detunings $Y(\delta_1, \delta_2)$. By symmetry, the power in the second fibre $Z(\delta_1, \delta_2)$ is the same as this with the axes $\delta_1\leftrightarrow \delta_2$ swapped. \textbf{(b)} Power response of both fibres as the detuning is scanned given the fixed relation $\delta_2=\delta_1+\pi/2$; see white dashed line in left figure. Result is obtained by numerically integrating the Ikeda map \eqref{eq: normalisedIkeda} without time $t$ dependence in the intra-fibre NLSE \eqref{eq: normalisedNLSE}. Here $\theta=2/15$, $A_{in} = 1$ and $\alpha_i=1$.}}  
\label{fig: numericalSS2}
\end{figure}

We first seek the time-independent steady states of the Ikeda map in both fibre loops. A pair of equations for the steady state fields is obtained by first integrating the NLSE Eq.~\eqref{eq: normalisedNLSE} over one round-trip, giving $A_n^{m}(z=1)= \exp(i {|A_n^m(z=0)|}^2 L_{\text{eff}} - \alpha_i/2)A_n^{m}(z=0)$. The dimensionless effective length $L_{\text{eff}} = (1-\exp(-\alpha_i))/\alpha_i$ accounts for the power decrease due to propagation losses, which reduces the accumulated nonlinear phase over a round-trip \cite{Hansson2015}.  Using this, we substitute for $A_n^{m}(1)$ in Eq.~\eqref{eq: normalisedIkeda}, imposing the constraint $A_n^{m+1}(0)=A_n^m(0)$.  A single equation for the two output powers $Y \equiv {|A_1|}^2$, $Z \equiv {|A_2|}^2$ can be obtained by eliminating the pump term $\sqrt{\theta} A_{in}$ and multiplying by the complex conjugate of both sides: 
\begin{multline} \label{eq: exactpumpindSS}
\left(1+ 2 e^{-\frac{1}{2}\alpha_i} \sqrt{1-\theta} \cos(Y  L_{\text{eff}}-\delta_1)  + e^{-\alpha_i}(1-\theta) \right) Y\\ = \left(1+ 2 e^{-\frac{1}{2}\alpha_i} \sqrt{1-\theta} \cos(Z  L_{\text{eff}}-\delta_2)  + e^{-\alpha_i}(1-\theta) \right) Z.
\end{multline}
To isolate a single resonance and obtain an approximated analytical solution of Eq. (\ref{eq: exactpumpindSS}), we expand the round-trip phases about the cavity anti-resonance: $ \exp(i (Y L_{\text{eff}}-\delta_1)) \rightarrow -1 - i(Y L_{\text{eff}}-\delta_1-\pi)$ and $\exp(i (Z  L_{\text{eff}}-\delta_2)) \rightarrow -1 - i(Z  L_{\text{eff}}-\delta_2-\pi)$. Expansion about the anti-resonance in fact corresponds to a resonance in both fibre loops separately, in the same manner as period-doubling instabilities in single-loop resonators \cite{Haelterman1992, Haelterman1992a, Haelterman1993}. We get:
\begin{multline} \label{eq: pumpindSS}
Y \left( {(1- \sqrt{1-\theta}e^{-\frac{1}{2}\alpha_i})}^2 + (1-\theta)e^{-\alpha_i}{(Y  L_{\text{eff}}-\delta_1-\pi)}^2 \right) \\
= Z \left( {(1- \sqrt{1-\theta}e^{-\frac{1}{2}\alpha_i})}^2 + (1-\theta)e^{-\alpha_i}{(Z  L_{\text{eff}}-\delta_2-\pi)}^2 \right)
\end{multline}
Solving Eq. (\ref{eq: pumpindSS}) for one of the powers in terms of the other gives three possible solutions for $Y(Z)$. Some example plots of these solutions with a fixed $\delta_2$ and various $\delta_1$ are shown in figure \ref{fig: TempSS}. In the case of a resonator with equivalent fibres, $\delta_1=\delta_2$, the powers follow a symmetric bistability curve identical to that in Fig. 1a) in \cite{Hill2020}. If the fibres are unequally detuned however,  the curve opens asymmetrically. If the detuning imbalance between the two fibres is relatively small, a separate closed orbit exists, close to which the solutions approximate an elliptic curve \cite{Washington2003}. These approximate local elliptic curves are indicated by red dashed boxes in figure \ref{fig: TempSS}. Although in the present paper we do not explore the implications of the existence of such curves in the stationary state diagram (a unique feature of our M\"{o}bius resonators), it is interesting to notice that elliptic curves are widely used in practical cryptography due to their amazingly rich group-theoretical structure that allows factoring integers efficiently \cite{Stein2009,Lenstra1987}. One could easily speculate that optical systems could be well-suited to study empirically open number-theoretical problems such as the famous Birch and Swinnerton-Dyer conjecture \cite{Silverman1992}.
Alternatively, one can resort to numerical methods to solve the exact equation for the steady state powers.
Numerically solving this equation (\ref{eq: exactpumpindSS}) using Newton's method shows a greatly extended (yet finite) range of solutions in the $Y, Z$ plane (figures \ref{fig: exactSS} and \ref{fig: exactSS2}). A large number of quasi-elliptic curve structures appear when asymmetric $Y \neq Z$ solutions are supported. However, solutions lying on repeated sub-structures at higher $Y$ and $Z$ are typically unstable and decay to the lowest branches .
Alternatively solutions can be plotted against the fibre loop detuning assuming a fixed pump power.  A distorted cavity resonance for in the intrafibre powers $(Y, Z)$ appears in the $(\delta_1, \delta_2)$ plane; as with the resonance in a standard Kerr cavity, this becomes broader as the round-trip losses are increased and more tilted for higher pump power or nonlinearity. The power in either fibre becomes asymmetric about $\delta_1=\delta_2$ however with otherwise indentical fibres the opposite asymmetry applies to power in the other fibre, that is $Y(\delta_1, \delta_2)=Z(\delta_2, \delta_1)$. An example is shown in figure \ref{fig: numericalSS2}. Scanning both detunings over the $2\pi$ domain while maintaining a constant offset between them allows the resonance to be crossed more than once, with different power dependencies in the two fibres depending on the choice of offset. 

\section{Dynamic Steady States \& M\"{o}bius Cavity Solitons}

\begin{figure}[h] 
\centering
\includegraphics[width=0.75\linewidth]{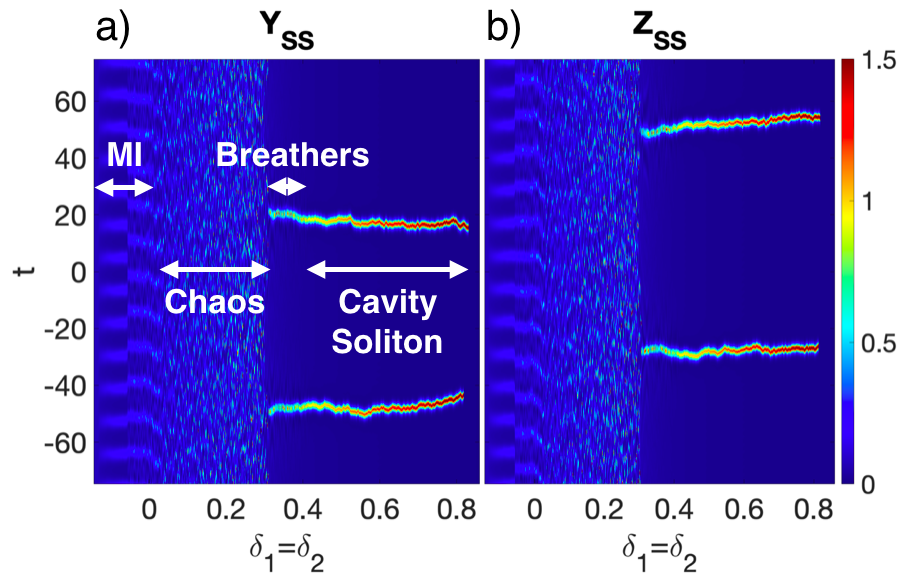}
\caption{\small{Stationary state power distributions $Y_{SS}$ in fibre 1 \textbf{(a)} and $Z_{SS}$ in fibre 2 \textbf{(b)} over time which emerge in a M\"{o}bius resonator composed of two identical fibres, as a function of common detuning from the pump $\delta_1=\delta_2$. The steady states are obtained by sequentially incrementing both $\delta_1$ and $\delta_2$ in steps of $0.01$, then propagating for $500$ round-trips of the dual Ikeda map Eq. \eqref{eq: normalisedIkeda}. Parameters: $A_{in} = 0.166$, $\beta_{3,n}=0$, $\eta_1=\eta_2=-1$, $\theta=2/15$, $\alpha_i=1/100$. Different regimes of dynamic steady-state behaviour are indicated with white labels and arrows; the steady state is time-independent for detunings outside these ranges.}}  
\label{fig: DynamSS}
\end{figure}

If the two fibres have identical parameters and detuning, we recover the typical regimes of ring resonator steady state behaviour, which are well described in the mean-field limit by a Lugiato-Lefever equation \cite{Coen2013a}. These are illustrated in figure \ref{fig: DynamSS}, which shows a sequence of homogeneous, modulation instability, chaotic, oscillating and stabilised cavity solitons followed finally by homogeneous steady states as the detuning of both fibres is gradually increased over the resonant interval.

\begin{figure}[h] 
\centering
\includegraphics[width=0.95\linewidth]{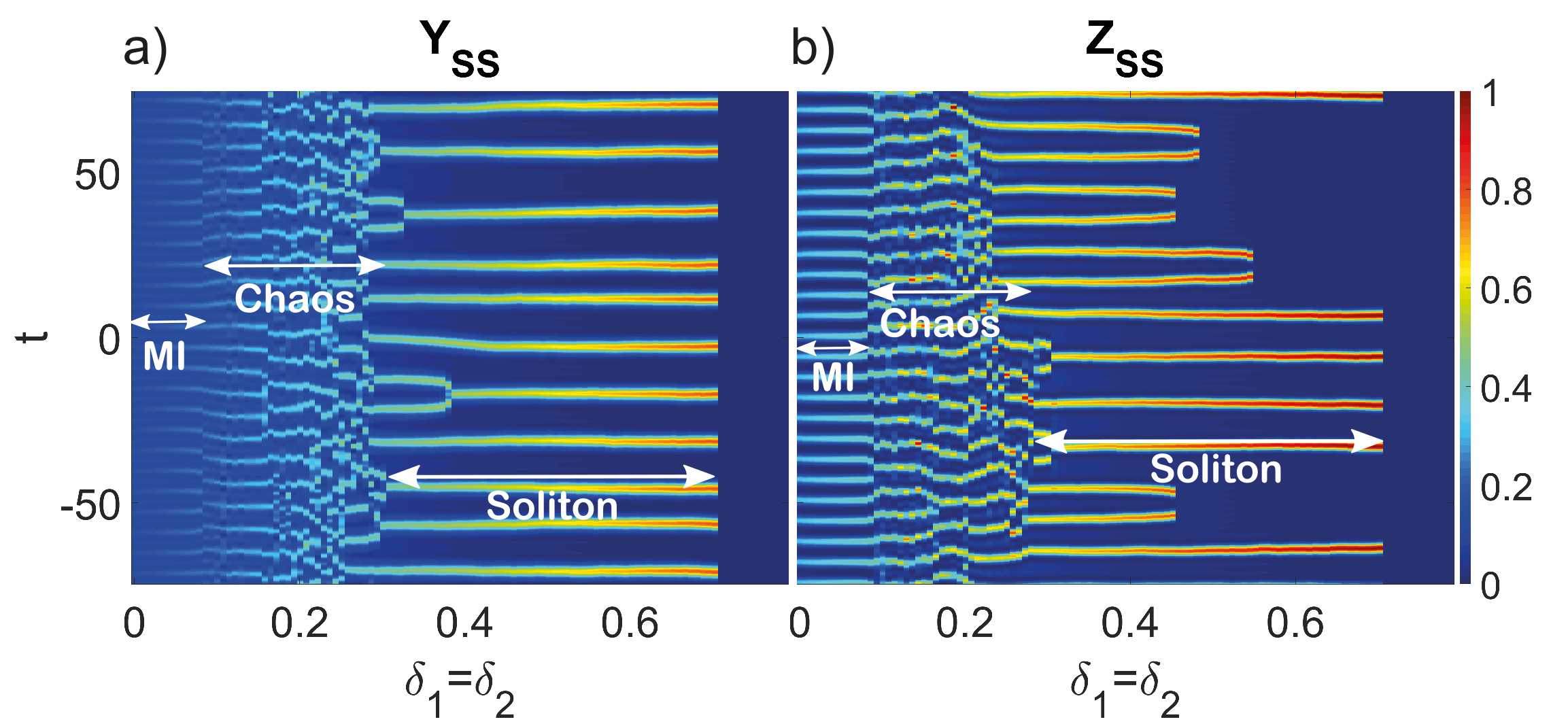}
\caption{\small{Stationary state power distributions $Y_{SS}$ in fibre 1 \textbf{(a)} and $Z_{SS}$ in fibre 2 \textbf{(b)} over time which emerge in a M\"{o}bius resonator composed of one anomalously-dispersive fibre $\eta_1=-1.5$ and a second, normally-dispersive fibre $\eta_2=1$, as a function of common detuning from the pump $\delta_1=\delta_2$. No third-order dispersion is present; parameters are $A_{in} = 0.166$, $\beta_{3,n}=0$, $\theta=2/15$, $\alpha_i=1/100$.}}  
\label{fig: DynamSS_UnbalacedB2}
\end{figure}

\begin{figure}[h] 
\centering
\includegraphics[width=0.95\linewidth]{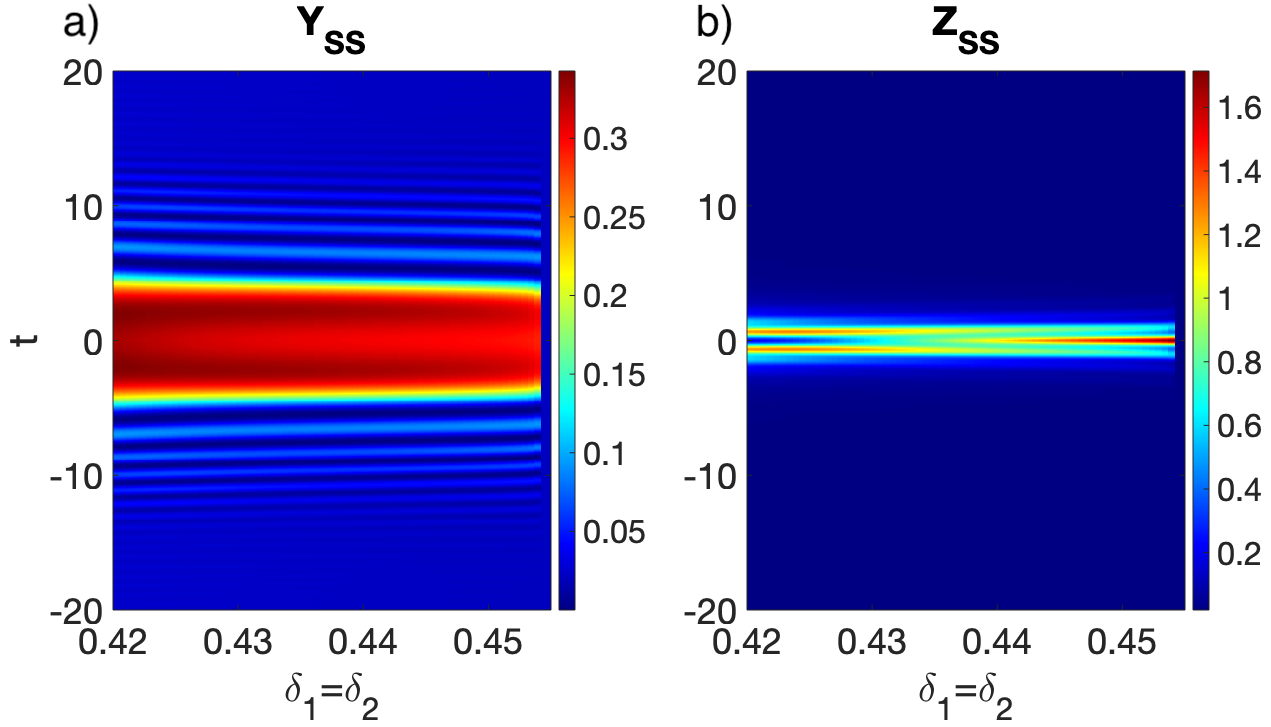}
\caption{\small{Stationary state power distributions $Y_{SS}$ in fibre 1 \textbf{(a)} and $Z_{SS}$ in fibre 2 \textbf{(b)} over time, which emerge in a M\"{o}bius resonator  composed of fibres with opposite group velocity dispersion (GVD) $\eta_1=-\eta_2=-1$ but otherwise identical parameters, as a function of common detuning from the pump $\delta_1=\delta_2$. }}  
\label{fig: DynamSS_OppGVD}
\end{figure}

\begin{figure}[h] 
\centering
\includegraphics[width=0.95\linewidth]{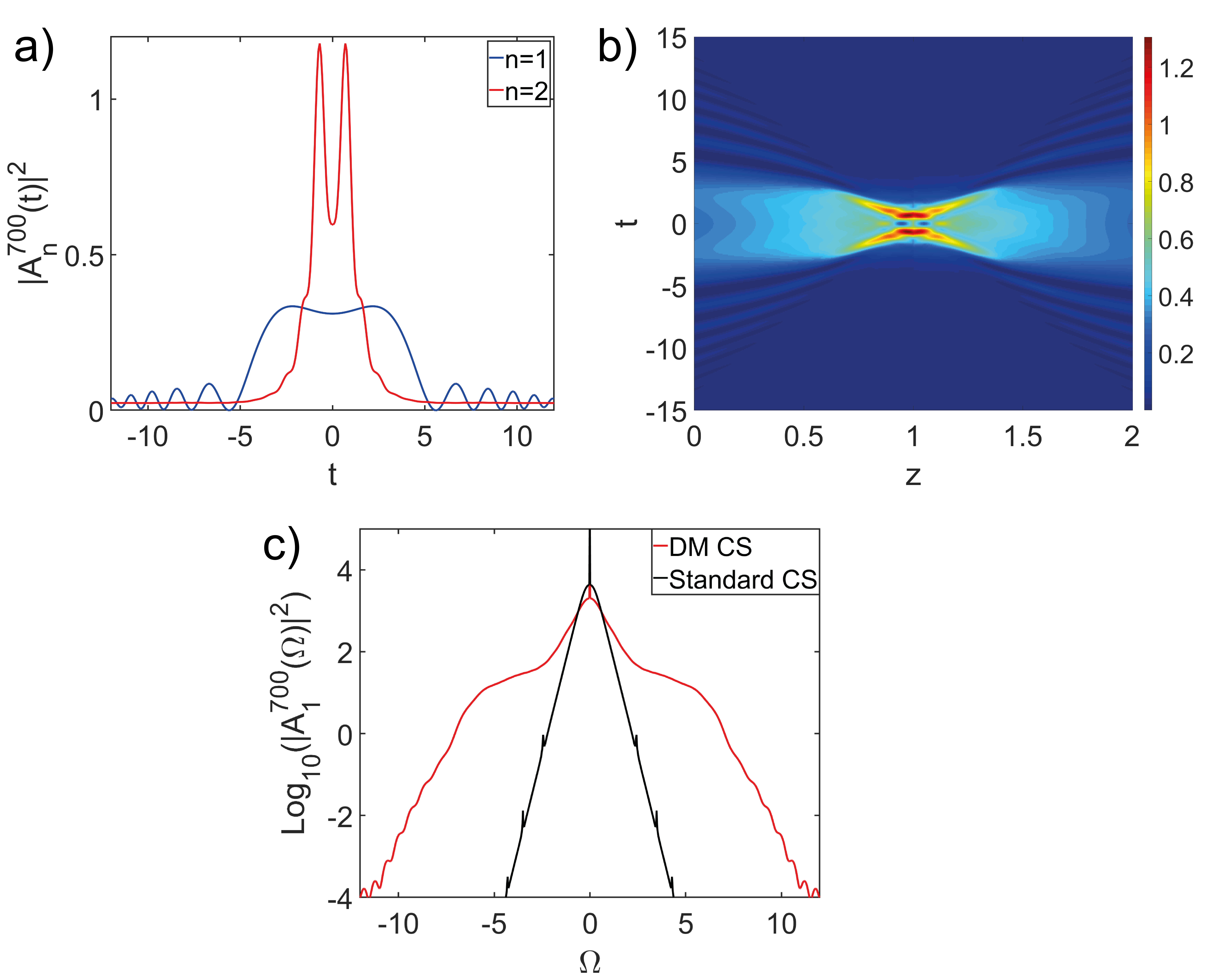}
\caption{\small{Example of the intracavity steady state power  in a M\"{o}bius resonator composed of two fibres with equal and opposite second-order dispersion $\eta_1=-\eta_2=-1$. Both fibres are detuned from the pump by $\delta_1=\delta_2=0.43$. \textbf{a)} Output power profiles from fibre one and fibre two after $700$ round-trips. \textbf{b)} Intra-fibre propagation of one field over two roundtrips, showing how the state continuously changes from the output distribution in one fibre to the other and back over one period of the M\"{o}bius cavity soliton (MCS). \textbf{c)} Comparison of the MCS spectrum with that of a standard Kerr cavity soliton which forms in an equivalent resonator with fibres that have the same dispersion, given the same pumping and initial condition as the dispersion oscillating resonator. The MCS bandwidth is considerably larger, particularly at lower powers.}  }
\label{fig: MobiusCS}
\end{figure}

\begin{figure}[h] 
\centering
\includegraphics[width=0.95\linewidth]{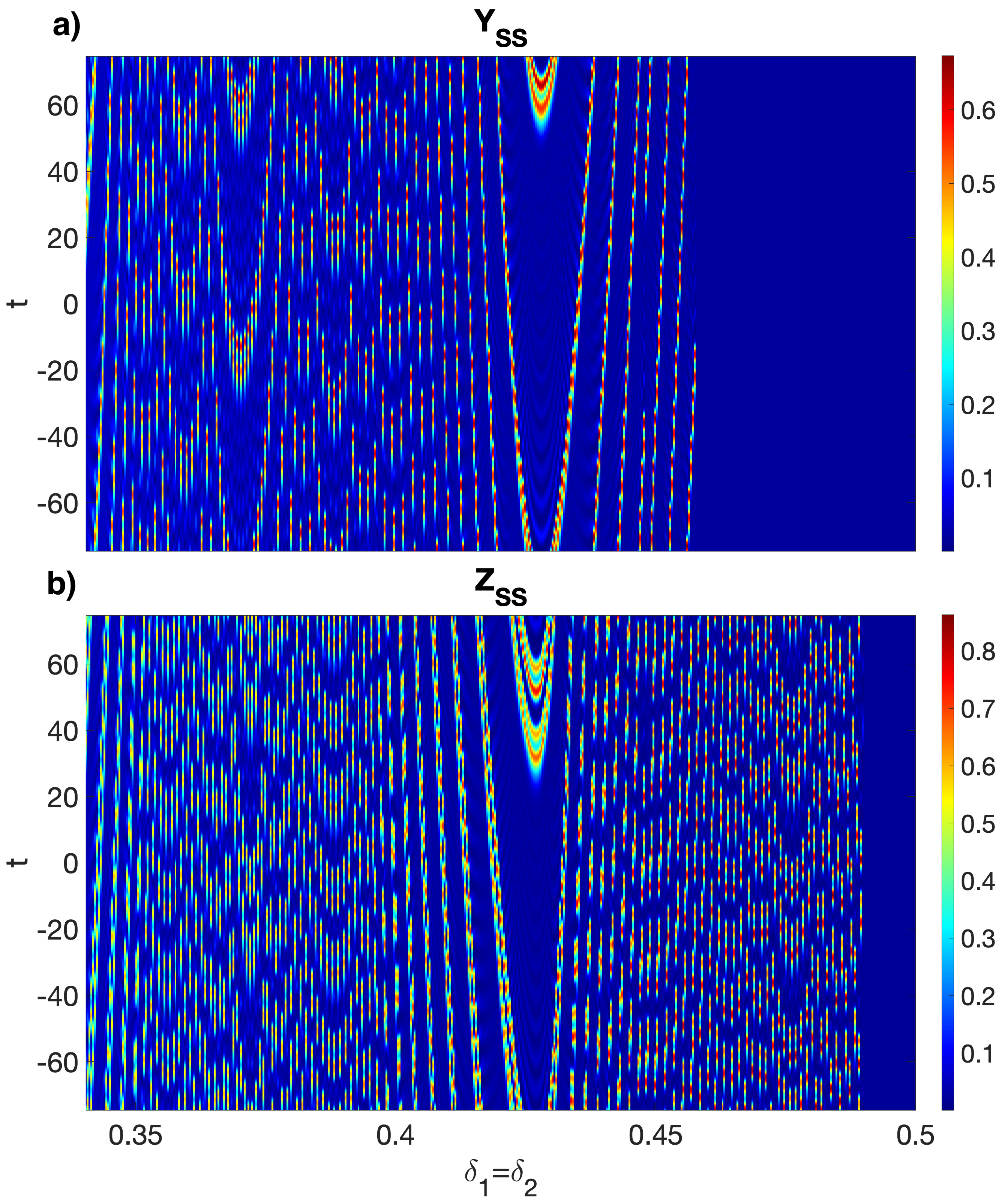}
\caption{\small{Intracavity power distributions over time \textbf{a)} $Y_{SS}$ in fibre one and \textbf{b)} $Z_{SS}$ in fibre two which emerges in a M\"{o}bius resonator composed of fibres with opposite second order $\eta_1=-\eta_2=-1$  but otherwise identical parameters, including third order $\beta_{3,1} =\beta_{3,2} = 2.6$ dispersion, as a function of common detuning from the pump $\delta_1=\delta_2$. A $t$-asymmetric M\"{o}bius cavity soliton appears for each detuning tested. Its central position with respect to $t$ shifts by a $\delta$-dependent amount after every $500$ round-trips, as higher-order dispersion imparts a frequency-dependent group velocity. }}
\label{fig: DynamSS_OppGVDandsameb3}
\end{figure}

\begin{figure}[h] 
\centering
\includegraphics[width=0.95\linewidth]{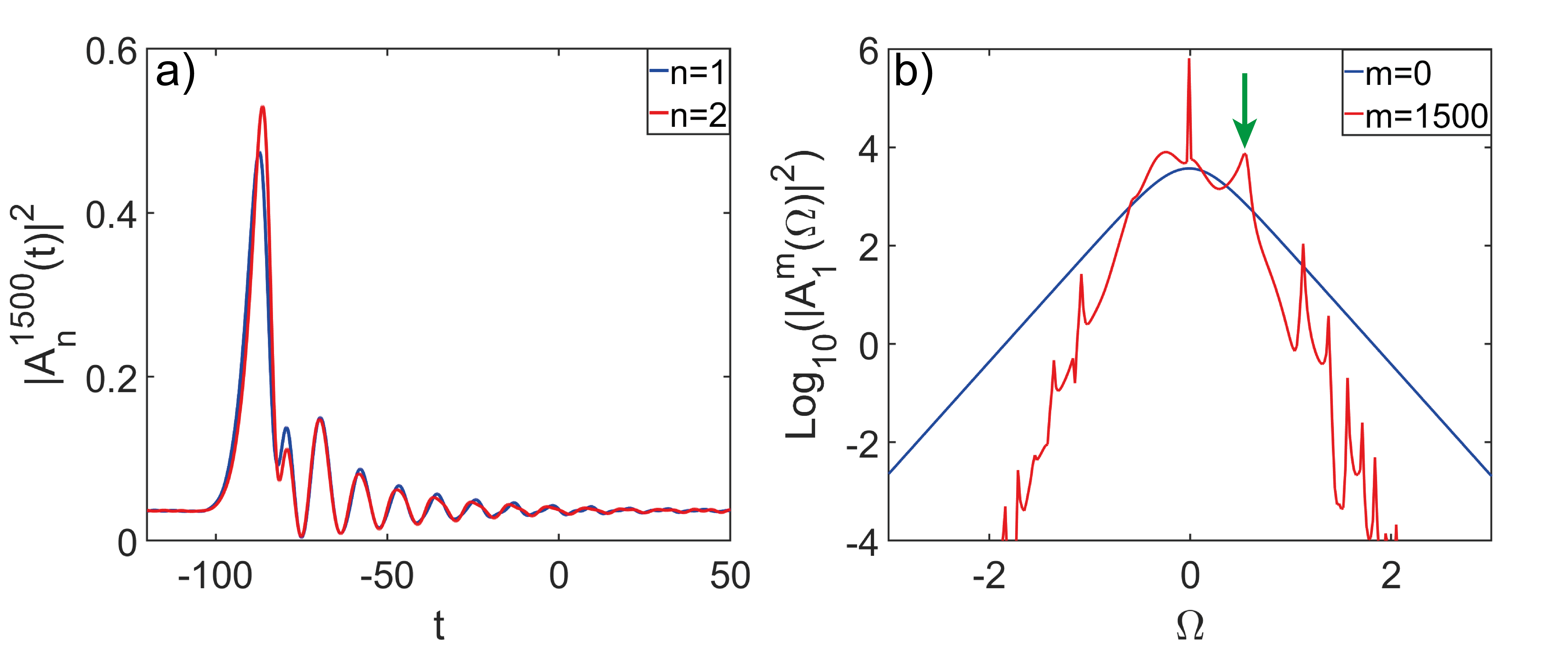}
\caption{\small{ \textbf{a)} Power profiles from fibre one and fibre two given $\delta=0.36$, $\eta_1=-\eta_2=-1$ and $\beta_{3,1} =\beta_{3,2} = 2.6$. \textbf{b)} Comparison of the output MCS spectrum with that of the input pulse. The green arrow indicates the resonant radiation peak.}  }
\label{fig: Mobius3rdorderCS}
\end{figure}

\begin{figure}[h] 
\centering
\includegraphics[width=0.95\linewidth]{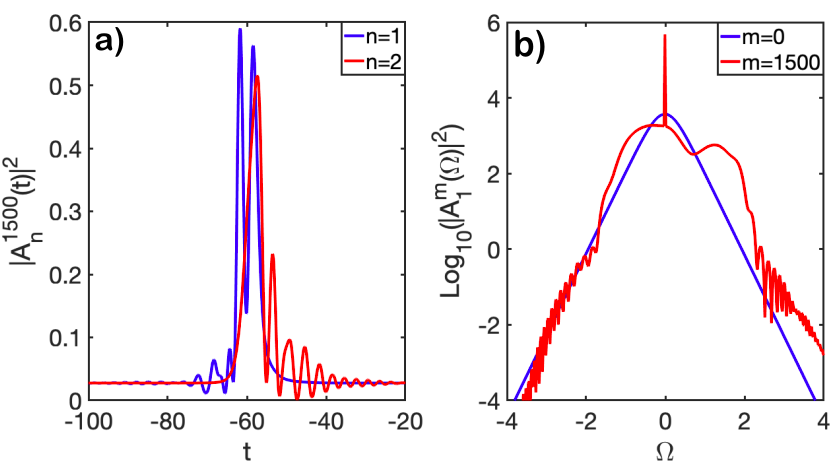}
\caption{\small{ \textbf{a)} Power profiles from fibre one and fibre two given $\delta=0.4$, $\eta_1=-\eta_2=-1$ and $\beta_{3,1} =-\beta_{3,2} = 2.6$. \textbf{b)} Comparison of the output MCS spectrum with that of the input pulse.}  }
\label{fig: Mobiusflipping3rdorderCS}
\end{figure}

\begin{figure}[h] 
\centering
\includegraphics[width=0.8\linewidth]{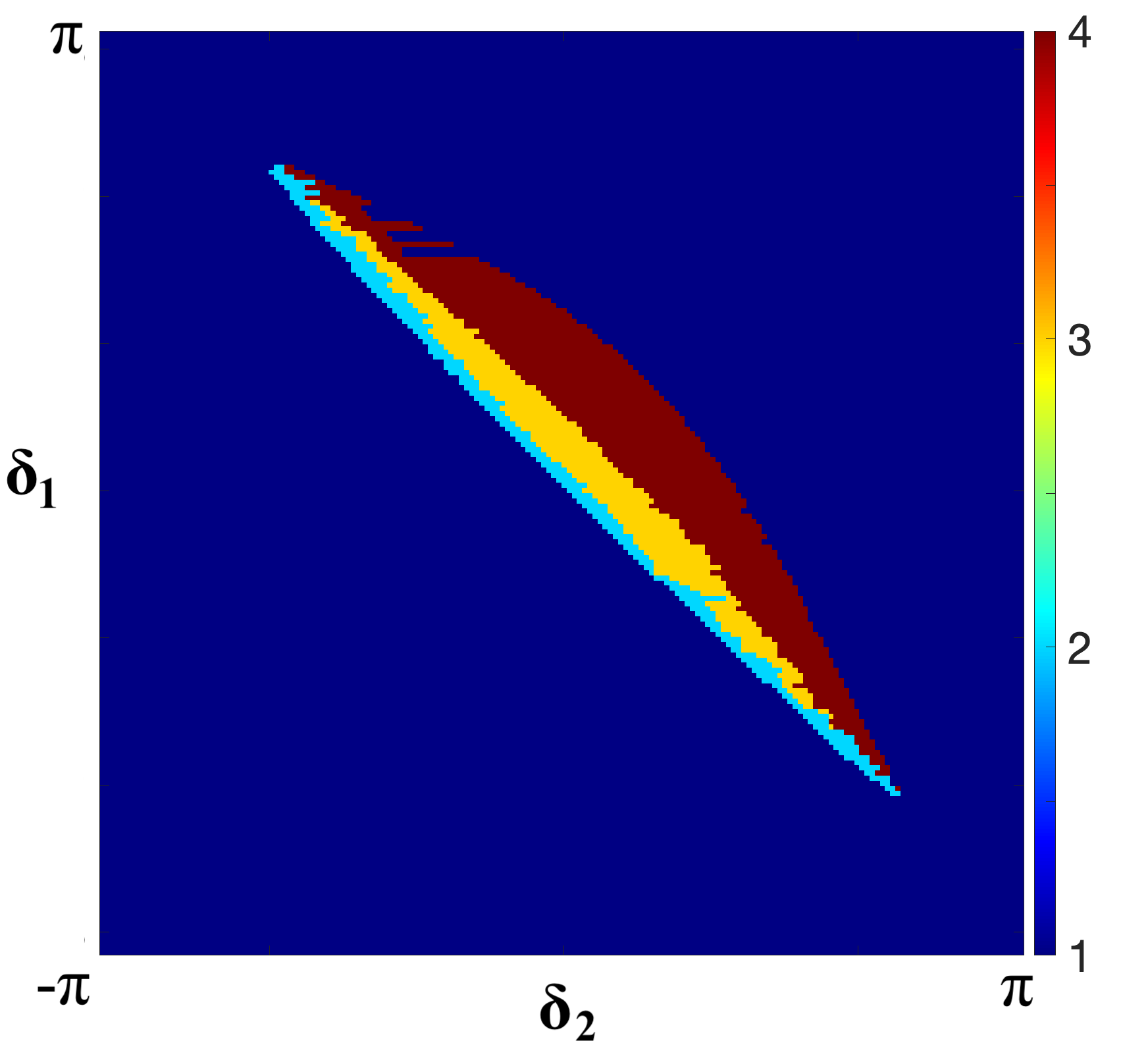}
\caption{\small{Dynamical steady state category in fibre 1 as a function of fibre detunings from resonance $(\delta_1, \delta_2)$. States are identified by a colour code; homogeneous states $1$,   modulation instability $2$, chaotic states $3$ and cavity solitons $4$.}}  
\label{fig: DynSS_2D}
\end{figure}

The two fibres within the loop may have completely different dispersive properties. In a first example we examine the steady states within a fibre resonator, where one fibre has normal dispersion $\eta_2=+1$ while the other has stronger anomalous dispersion $\eta_1=-1.5$. Surprisingly similar behaviour regimes to those seen with identical anomalous fibres emerge, with clear transitions from stationary MI patterns to chaos followed by cavity soliton formation (figure \ref{fig: DynamSS_UnbalacedB2}). When multiple solitons form collisions may occur; at lower detuning this results in the solitons merging, whereas they tend to collapse for higher detuning. Reducing the strength of the anomalous dispersion in the second fibre decreases the detuning ranges for which MI and cavity solitons are supported. 
Considering a pair of fibres which have equal and opposite dispersion, $\eta_1=-\eta_2=-1$, neither modulation instability nor chaotic steady states emerge. However, instead of the usual Kerr cavity soliton, a new kind of compact state which we term \textit{M\"{o}bius cavity soliton} (MCS) appears with a complicated yet stable waveform at the output of both fibres. In all that follows, the output states given varying values of detuning $\delta_1=\delta_2$ are recorded after propagating the initial condition over many round-trips of the dual Ikeda map Eq. \eqref{eq: normalisedIkeda}. Specifically, at each step we increment $\delta_1=\delta_2$  by a small amount and propagate over $500$ round-trips, using the output from the previous $\delta_1=\delta_2$  increment as the initial condition. This is typically sufficient to obtain convergence to the steady state in each case.  At the first integration step we choose the initial condition $A_n(t) = 1.2 \sech(1.2 t)$ for $n=1,2$.  We also choose $A_{in} = 0.166$, $\theta=2/15$ and $\alpha_i=1/100$. Figure \ref{fig: DynamSS_OppGVD} plots the output power from both fibres as a function of detuning. Under this constraint, the MCS appears only within a small detuning range $\delta \in (0.42, 0.455)$, and its structure changes somewhat within this range. Outside of this range only homgeneous steady states appear, with the modulation instability and chaotic regimes apparently suppresed.
 On closer examination, plotting the field's intra-fibre evolution over two round-trips reveals that the stationary fibre outputs are snapshots of a stable periodic state which oscillates smoothly between the two (figure \ref{fig: MobiusCS}). The periodic behaviour arises as the field experiences cyclic second-order dispersion as it travels through both loops of the resonator; a similar state was previously found in a dispersion-modulated fibre ring \cite{Gavrielides2004}. The spectrum of this MCS is considerably wider than that of the standard cavity soliton given the same pump power and detuning, and therefore is a promising seed for a broadband frequency comb when emitted in a pulse train.
Including third order dispersion $\beta_{3, n} = 2.6$ in both fibres, the MCS output becomes asymmetric in time and acquires a finite group velocity. It is stable for an extended detuning range $\delta \in (0.34, 0.45)$, however unlike the previous MCS which results from $\beta_{3, n} = 0$ the two fields within the resonator are distinct and do not periodically transform into each other. As a consequence the output from either fibre swaps each round-trip (but appears stationary when examined every second round-trip). From this it is clear that the initial condition affects steady state stability, as the field which first travelled through the anomalously-dispersive fibre supports a MCS to a slightly larger detuning limit $\delta = 0.49$ than that which started in the fibre with normal dispersion, $\delta = 0.46$. See figure \ref{fig: DynamSS_OppGVDandsameb3}. This MCS has a highly oscillatory structure in time and a narrower, peaked spectrum with resonant radiation resulting from the third-order dispersion clearly visible (figure \ref{fig: Mobius3rdorderCS}).
If the sign of $\beta_{3, n}$ swaps sign between fibres as well $\eta_n$, a similar MCS forms in both fibre loops. 
The oscillating $\beta_{3, n}$ leads to a reduced MCS group velocity compared to the state formed with a common $\beta_{3, n}$ as well as a reversed $t$-asymmetry between the power outputs (figure \ref{fig: Mobiusflipping3rdorderCS}). Further, the resonant radiation spectral peaks do not appear since consistent phase-matching cannot be achieved under these conditions.
%
Allowing for unequal detuning from resonance in the two fibres $\delta_1 \neq \delta_2$ adds an additional dimension to the resonator's parameter space and considerably extends the possible  existence of dynamical steady states. Performing a similar survey of time dependent steady states over $(\delta_1, \delta_2) \in [-\pi, \pi] \times [-\pi, \pi]$ as was done in figure \ref{fig: DynamSS} for a resonator with $\eta_1=-1.5$, $\eta_2=1$ yields map as shown in figure \ref{fig: DynSS_2D}. Time dependent states are concentrated around the joint resonance close to the line $\delta_1+\delta_2=0$. Proceeding from the the negative $(-\pi,-\pi)$ to positive $(\pi,\pi)$ we find the familiar sequence of homogeneous, modulationally unstable, chaotic, solitonic and homogeneous states. These state domains show some curvature in the $(\delta_1, \delta_2)$ plane, which we expect to increase with pump power and round-trip losses as the resonance assuming homogeneous states in figure \ref{fig: numericalSS2} does. 
We note that the dynamical steady-states are dependent on the sequence of parameters tested and their intial values. For example, if the scan of increasing detuning shown in figure \ref{fig: DynamSS_OppGVD} had started with $\delta_1=\delta_2=0.4$, the corresponding steady state would be homogeneous. On increasing  $\delta_1=\delta_2$ from this point, we would not observe the formation of M\"{o}bius solitons in the same detuning range as in \ref{fig: DynamSS_OppGVD} unless an external perturbation was added. It appears that the chaotic MI phase from which cavity solitons typically emerge is supressed in resonators with $\eta_1=-\eta_2=-1$,  meaning they will not appear spontaneously on increasing detuning from the homogeneous state. Likewise, had we examined the same detuning ranges in reverse order a different sequence of states would result. The analysis of bifurcation, hysteresis and multi-stability in single fibre resonators is a complex area of study in itself \cite{ParraRivas2016, ParraRivas2017, ParraRivas2018}, so we propose a more thorough examination of these as applied to the M\"{o}bius resonator in future works.

\section{Modulation Instability}

\begin{figure}[h] 
\centering
\includegraphics[width=0.95\linewidth]{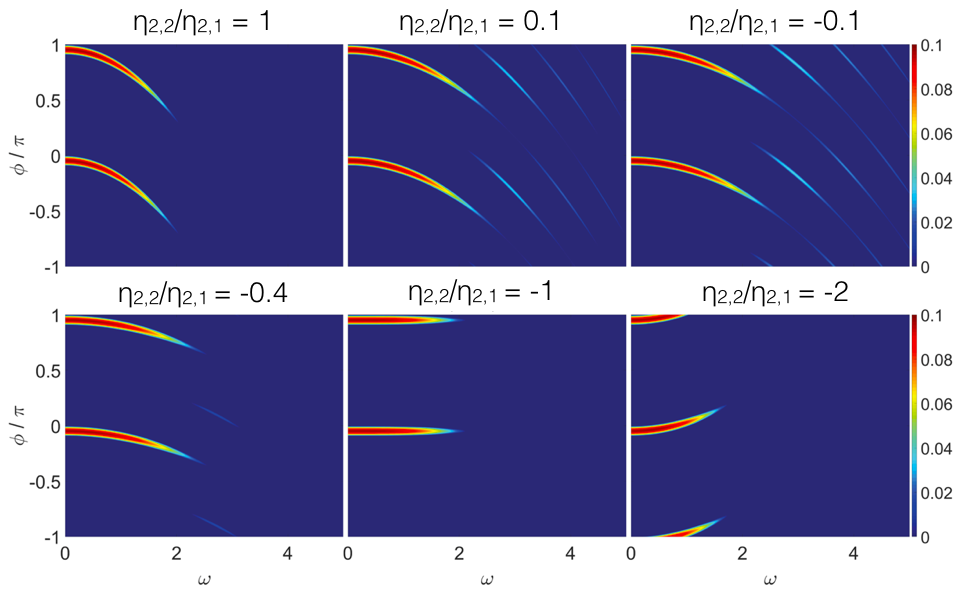}
\caption{\small{Modulation instability gain as a function of frequency $\omega$ and common round-trip phase $\phi_1=\phi_2=\phi$ for various relative second-order dispersions in the two resonator fibres. }  }
\label{fig: MobiusMIgain}
\end{figure}

\begin{figure}[h] 
\centering
\includegraphics[width=0.95\linewidth]{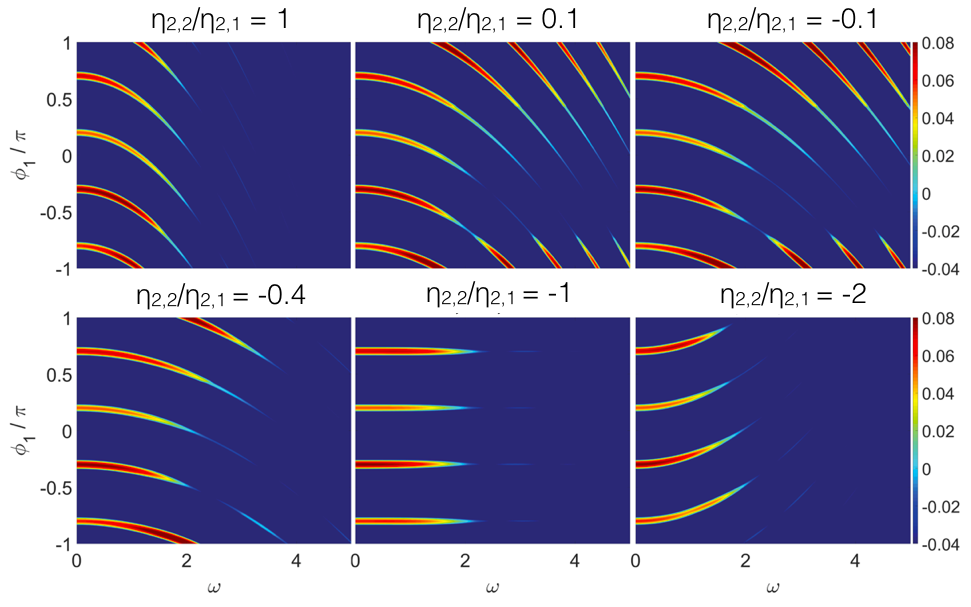}
\caption{\small{Modulation instability gain as a function of frequency $\omega$ and round-trip phase in the first fibre $\phi_1=\phi_2-\pi/2$ for various relative second-order dispersions in the two resonator fibres. }  }
\label{fig: MobiusMIgain_PiOver2Shift}
\end{figure}

\begin{figure}[h] 
\centering
\includegraphics[width=0.95\linewidth]{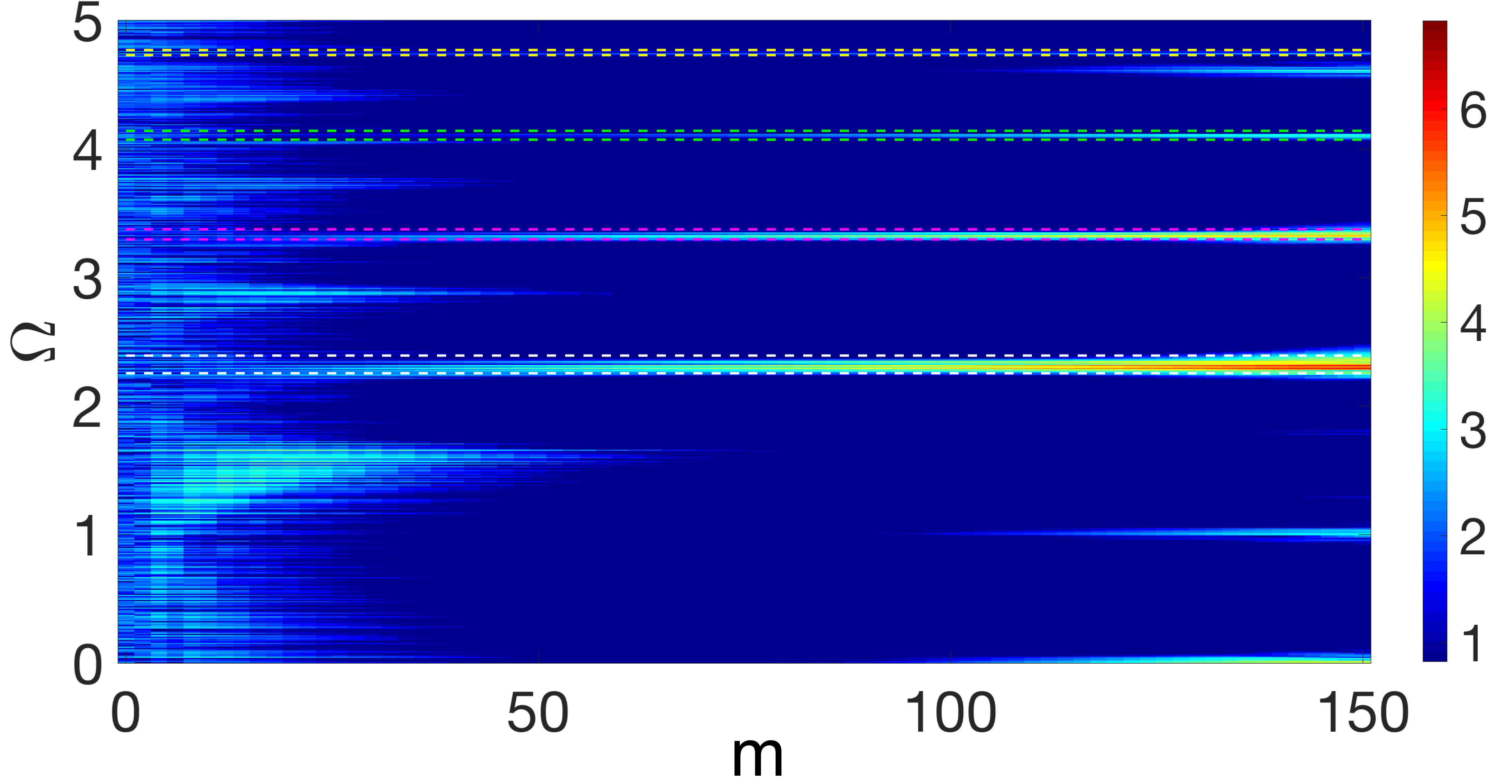}
\caption{\small{Modulation instability spectrum that emerges over $m$ round-trips in a M\"{o}bius resonator with $\delta_1=\delta_2=0.6\pi$, $P_1=P_2=0.2$, $\eta_{2, 2} =  10\eta_{2, 1} =1$. Colour axis shows the intracavity spectral power of the second fibre on a logarithmic scale, as measured every second round-trip. Only the positive half of the spectrum is shown since it is symmetric about $\Omega=0$. The spectrum in the first fibre loop is nearly identical and features sidebands with the same strength and position as those shown here. Dashed white, magenta, green and yellow lines indicate the band edges as predicted by the instability gain spectrum resulting from Floquet analysis.}  }
\label{fig: MobiusMI_numerics}
\end{figure}

\begin{figure}[h] 
\centering
\includegraphics[width=0.95\linewidth]{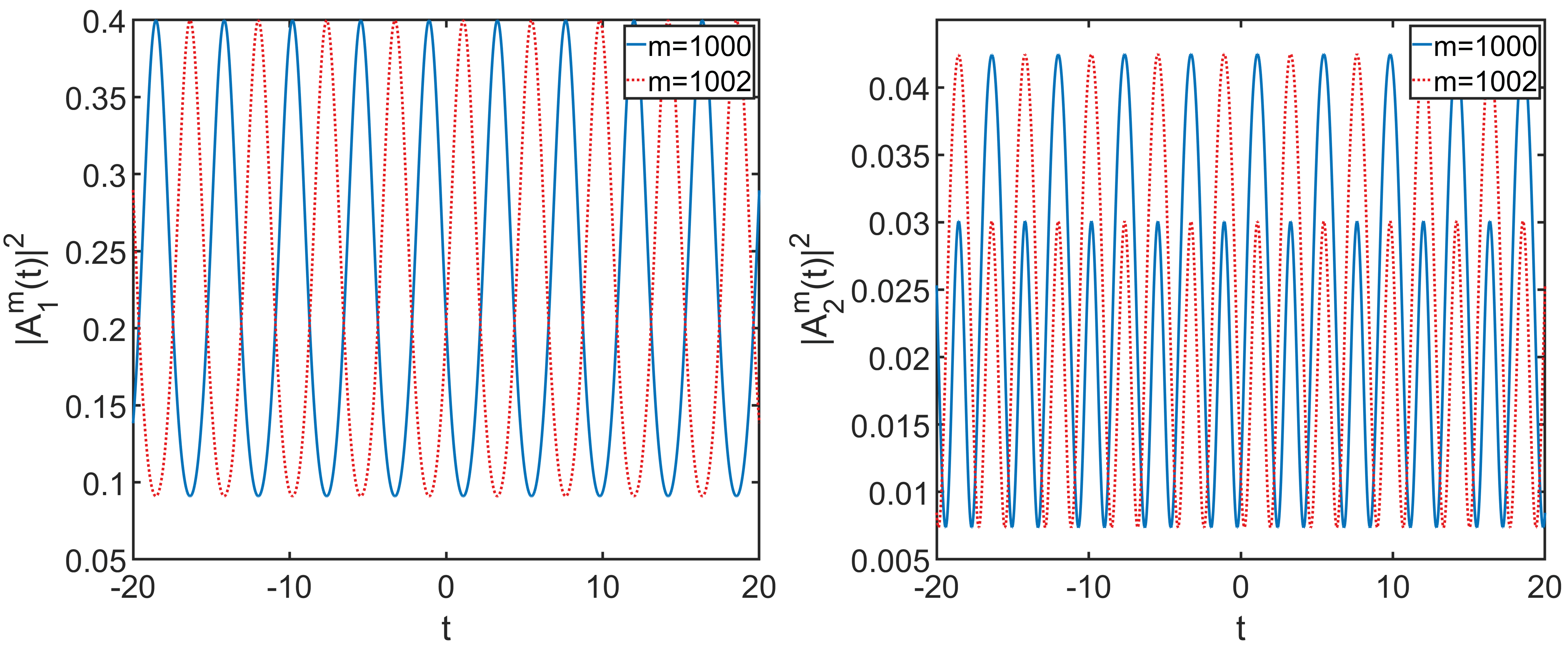}
\caption{\small{Intracavity power over time in fibre one (a) and fibre two (b) after $m=1000, 1002$ round-trips. Here the eigenvalues of the Floquet matrix $W$ are imaginary, so  period doubling manifests as a repetition of the Faraday MI pattern within each fibre every two round-trips, with the phase of the pattern reversing each time (hence four round-trips are necessary for a complete cycle $A_n^{m}(t)=A_n^{m+4}(t)$). Parameters are chosen such that a single Faraday (P2) instability branch is significantly excited over this many round-trips; $\delta_2=\pi/2$, $\delta_1=\pi$, ${|A_{in}|}^2=1.79$, $\eta_{2, 1} =  10\eta_{2, 2} =1$. }  }
\label{fig: MobiusMI_PeriodDoubling}
\end{figure}

\begin{figure}[h] 
\centering
\includegraphics[width=0.95\linewidth]{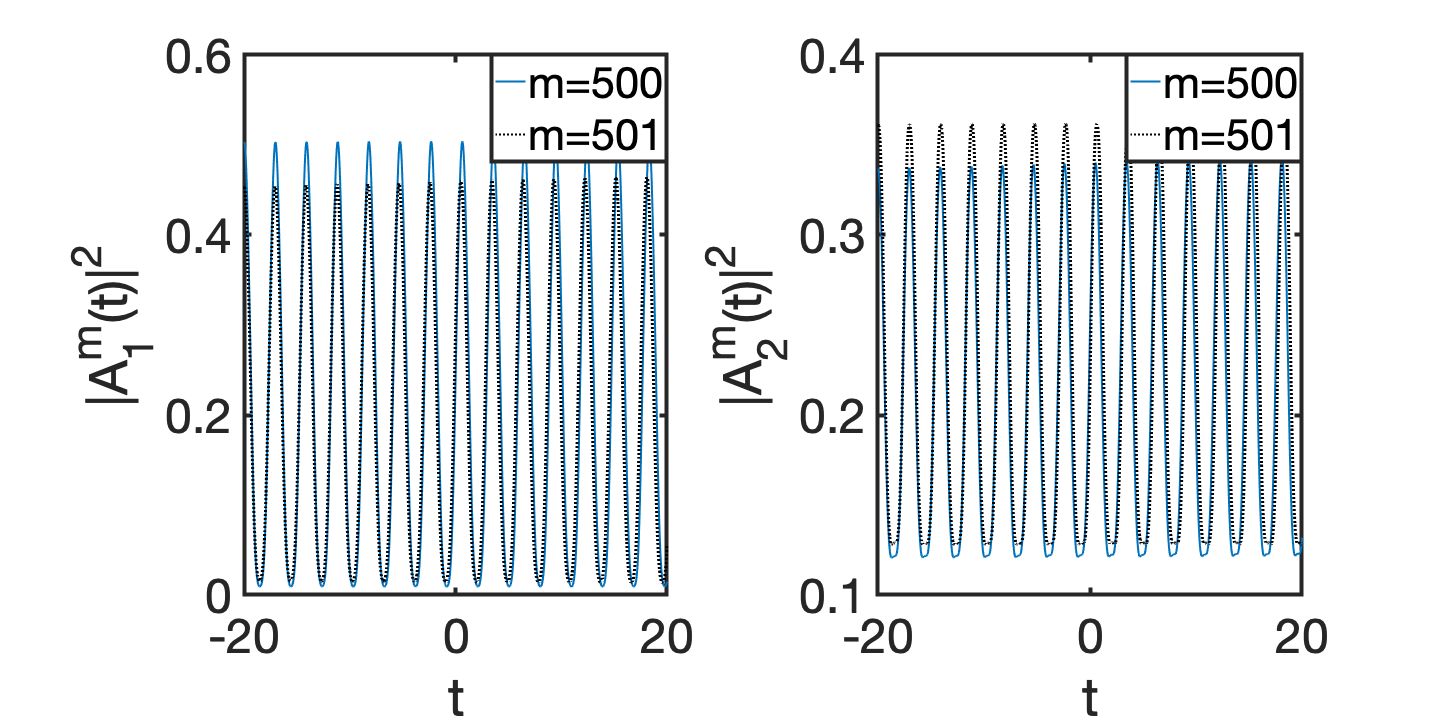}
\caption{\small{Intracavity power over time in fibre one (a) and fibre two (b) after $m=1000, 1002$ round-trips. Here the eigenvalues of the Floquet matrix $W$ are real, so the MI pattern repeats exactly in both fibres every second round-trip, \textit{i.e.} $A_n^{m}(t)=A_n^{m+2}(t)$. The intracavity pattern varies slightly every other round trip, so $A_n^{m}(t) \neq A_n^{m+1}(t)$. Parameters are chosen such that a single P1 instability branch is significantly excited over this many round-trips; $\delta_2=\pi/2=\delta_1$, ${|A_{in}|}^2=1.79$, $\eta_{2, 1} =  10\eta_{2, 2} =1$. }  }
\label{fig: MobiusMI_StandardPeriod}
\end{figure}

The presence of two fibres in the resonator with distinct dispersive properties raises the possibility of strongly modified modulation instability compared with that observed in homogeneous fibre resonators \cite{Hansson2015,Haelterman1992b,Coen1997}. The step-wise dispersion modulation seen by light over a complete round-trip of the resonator might be expected to result in similar instabilities as a resonator composed of a single fibre with oscillating dispersion, as presented in several previous works \cite{Conforti2014, Conforti2016, Copie2016,Bessin2019}. To investigate this, we adapt the Floquet analysis of \cite{Conforti2016} to the M\"{o}bius resonator. The only significant extension required is that perturbations in both fibres must be monitored simultaneously to describe a complete round-trip, meaning we have a system of four simultaneous equations to solve rather than two. To facilitate analysis, we  incorporate small round-trip losses into the boundary condition of the Ikeda map Eq. (\ref{eq: normalisedIkeda}), which now reads:
\begin{equation} \label{eq: IkedaMI}
\begin{split}
A_1^{m+1}(z=0, t) = \sqrt{\theta} A_{in} + \sqrt{\rho} e^{-i \delta_2} A_2^m(z=1,t)\\
A_2^{m+1}(z=0, t) = \sqrt{\theta}  A_{in} + \sqrt{\rho}  e^{-i\delta_1} A_1^m(z=1, t)\\
\end{split}
\end{equation}
\begin{equation} \label{eq: cavityNLSEMI}
\partial_z A_n^m = -i\eta_n {\partial_t}^2 A_n^m + i {|A_n^m|}^2 A_n^m
\end{equation}
with $\rho+\theta<1$. We write the total field in the two resonators as $A_1^m(z,t) = \overline{A}_1 + \check{a}^m(z, t) +i \check{b}^m(z, t)$,  $A_2^m(z,t) = \overline{A}_2 + \check{c}^m(z, t) +i \check{d}^m(z, t)$. $\check a, \check b,  \check c, \check d$ are real perturbations and the time-independent stationary state in resonator $n$ is
\begin{equation} \label{eq: MItimeindSS}
\overline{A}_n(z) = \sqrt{P_n} \exp\left(i P_n z + \psi_n \right).
\end{equation}
As discussed earlier, there is no analytic solution available for the steady state powers $P_n$ and phases $\psi_n$, though it is straightforward to find them numerically by integrating a time-independent version of the Ikeda map. Considering solely evolution through the first fibre loop, the linearised NLSE for the perturbation coefficients reduces to a $2\times2$ coupled pair of ordinary differential equations,
\begin{equation} \label{eq: linearODE1}
\frac{d}{dz} \begin{pmatrix}
a^m(z, \omega) \\ 
b^m(z, \omega) 
\end{pmatrix} 
= 
\begin{pmatrix}
0  &  -\eta_1\omega^2 \\ 
\eta_1 \omega^2+ 2 P_1  & 0
\end{pmatrix} 
\begin{pmatrix}
a^m(z, \omega) \\ 
b^m(z, \omega) 
\end{pmatrix} .
\end{equation}
Note the perturbations have been Fourier transformed into the frequency ($\omega$) domain. This problem has solutions with eigenvalues 
\begin{equation} \label{eq: eigval}
\pm k_1= \pm \sqrt{\eta_1 \omega^2\left( \eta_1 \omega^2+ 2 P_1 \right) }
\end{equation}
and corresponding eigenvectors
\begin{equation} \label{eq: eigvec}
\mathbf{v}_{+, -}= \begin{pmatrix}
\cos(k_1 z)\\ 
\frac{k_1}{\eta_1\omega^2} \sin(k_1 z)
\end{pmatrix}, 
\begin{pmatrix}
-\frac{\eta_1\omega^2}{k_1} \sin(k_1 z)\\ 
\cos(k_1 z)
\end{pmatrix} . 
\end{equation}
The eigenvectors define a fundamental solution matrix for the ODE (\ref{eq: linearODE1}) $X_1(z) = ( \mathbf{v}_+ , \mathbf{v}_- )$. A completely analogous solution matrix will exist for the perturbations in the second fibre; 
\begin{equation} \label{eq: FSS}
X_2(z) = \begin{pmatrix}
\cos(k_2 z)  &  -\frac{\eta_2 \omega^2}{k_2} \sin(k_2 z)\\ 
\frac{k_2}{\eta_2 \omega^2} \sin(k_2 z) & \cos(k_2 z)
\end{pmatrix}, 
\end{equation}
These can be combined into a single 4x4 block diagonal matrix which describes the evolution of all perturbations in the $\mathbf{a}^m(z)=(a^m, b^m, c^m, d^m)$ basis as
\begin{equation} \label{eq: masterFSS}
X(z_1, z_2) = \left[ \begin{array}{c|c}
X_1 & O_2  \\ \hline
O_2  & X_2 \\
\end{array} \right] \\
\end{equation}
given $O_2$ is the 2x2 null matrix. Meanwhile the boundary conditions can be implemented by a combined rotation
\begin{equation} \label{eq: masterBC}
\Gamma = \sqrt{\rho}\left[ \begin{array}{c|c}
O_2 & \Gamma_2  \\ \hline
\Gamma_1  & O_2 \\
\end{array} \right] \\
\end{equation}
where
\begin{align} \label{eq: BC}
\Gamma_n &= \begin{pmatrix}
\cos(\phi_n) & -\sin(\phi_n)   \\
\sin(\phi_n) & \cos(\phi_n)
\end{pmatrix} 
\end{align}
and $\phi_1= P_1-\delta_1   + \psi_1-\psi_2$, $\phi_2=P_2 -\delta_2 + \psi_2-\psi_1$. The combined matrix describing a complete round-trip evolution is the product of these, $W = \Gamma X(1, 1)$. 
The eigenvalues and eigenvectors of $W$ can be solved for analytically, however the expressions are not particularly tractable. The four eigenvalues consist of two pairs with equal magnitudes but opposite sign, $\pm \lambda_1, \pm \lambda_2$; since the eigenvalue modulus is what determines instability gain, we need only consider one of each pair. The gain for a particular frequency $\omega$ is then \cite{Conforti2016}
\begin{equation} \label{eq: MobiusMIgain}
g(\omega)=\log\left(\max({|\lambda_1(\omega)|, |\lambda_2(\omega)|})\right)
\end{equation}
We plot the gain assuming different values of dispersion in both fibres in figure \ref{fig: MobiusMIgain}, assuming the detuning in both fibres is the same. At least two instability branches are always present at low frequencies. Increasing the GVD value in the first fibre $\eta_1$ relative to the fixed GVD in the second fibre $\eta_2=1$ alters the curvature of the instability branches; it also gives rise two additional branches at higher frequencies when the difference in dispersion magnitude between the two fibres is sufficiently big. This is consistent with the findings of \cite{Conforti2016}. 
Here we have an additional degree of freedom in the relative detuning between the two fibres, which can modify the both the extent and positions of the branches. For example, choosing $\phi_2=\phi_1+\pi/2$ leads to the modified gain in figure \ref{fig: MobiusMIgain_PiOver2Shift}, which shows the branches enabled by dispersion oscillation in figure \ref{fig: MobiusMIgain} become significant at low frequencies for any relative dispersion $\eta_{1} / \eta_{2}$. 
It is interesting to note that when the two loops have dispersion with equal magnitude but opposite signs, the instability branches become flat. This means that for each steady state parametrized by the phase $\phi$, either we have no gain, either the gain is peaked around $\omega=0$. Hence the steady state is either stable or unstable with respect to zero frequency perturbations, that is the unstable state of a multi-stable response. This fact explains why no modulation instability is observed for opposite dispersions, as we found in the previous section (see Fig. \ref{fig: DynamSS_OppGVD}).
To check the accuracy of the Floquet analysis' predictions we numerically simulate MI in a M\"{o}bius resonator with $\delta_1=\delta_2=0.6\pi$, $P_1=P_2=0.2$, $\eta_{2, 1} =  10\eta_{2, 2} =1$. 
The Floquet theory indicates that to first order there should be four pairs of instability bands under these conditions, and indeed the numerically integrated spectrum shows three sidebands developing after $100$ round-trips (see figure \ref{fig: MobiusMI_numerics}).
Instability branches may vary in their period behaviour; MI patterns on certain branches repeat exactly after every two round-trips, whereas those on others repeat every four round-trips. The period-doubling behaviour associated with Faraday-type instability branches is well documented in other works \cite{Hansson2015, Conforti2016, Bessin2019}; an example is presented in figure \ref{fig: MobiusMI_PeriodDoubling}, in which the intracavity power in either fibre shows the same MI pattern after two round-trips, but exactly out-of-phase with respect to the original pattern. The power time series in both fibres repeats exactly every four round-trips, corresponding to a full period of the Faraday instability. This contrasts with homogeneous or dispersion oscillating cavities, where the Faraday instability gives rise to a period two pattern. This periodicity is explained by the fact that the unstable eigenvalues of the Floquet matrix $W$ are purely imaginary over these branches, i.e. they have a phase of $\pm\pi/2$. The perturbations are again in-phase after four round-trips, which eventually generates the observed sequence. For comparison an instability pattern that develops with real Floquet eigenvalues, which repeats exactly every two round-trips, is presented in figure \ref{fig: MobiusMI_StandardPeriod}. the A detailed analysis of parametric instabilities of the M\"obius resonator and they period doubling is outside the scope of this paper and it will be reported elsewhere.

\section{Conclusion}
We have found new dissipative structures in a M\"{o}bius optical fibre resonator, which to our knowledge has not been studied previously. When continuous-wave solutions are modulationally stable, their powers define unusual bistability curves which may be approximated by elliptic curves in certain limits. A variety of time-dependent localised and periodic states which cannot be realised by standard fibre resonators are supported, including exotic cavity solitons and extended modulation instability. These are enabled by the ability to tune the resonances and dispersive properties of both fibres in the resonator independently. We anticipate that M\"{o}bius cavity solitons will be of interest to researchers working on frequency comb generation, owing to their broadened spectrum compared to the typical Kerr cavity soliton. 

\section*{Acknowledgments}
C.M. acknowledges studentship funding from EPSRC under CM-CDT Grant No. EP/L015110/1. F.B. acknowledges support from the German Max Planck Society for the Advancement of Science (MPG), in particular the IMPP partnership between Scottish Universities and MPG.

\bibliography{Moebius}

\end{document}